\documentstyle[12pt]{article}

\title{On discrete symmetry for spin 1/2 and spin 1 particles in external
monopole field  and quantum-mechanical property of self-conjugacy}
\date{}
\medskip

\author {V.M.Red'kov \\
     Institute of Physics, Belarus Academy of Sciences\\
F Skoryna  Avenue 68, Minsk 72, Republic ob Belarus\\
e-mail: redkov@dragon.bas-net.by}

\begin{document}
\maketitle


\begin{abstract}
Particles of spin 1/2 and 1 in external Abelian
monopole field are considered.  $P$-inversion-like operators
$\hat{N}$, commuting
with the~respective Hamiltonians, are constructed:
$\hat{N}_{bisp.}$ is diagonalized onto
 the~relevant wave functions, whereas $\hat{N}_{vect.}$ does not.
Such  a~paradox is rationalized through noting that both
these operators are not self-conjugate.
It is shown that  any $N$-parity selection rules cannot be produced.
Non-Abelian problems for doublets of spin 1/2 and 1 particles are
considered; corresponding discrete operators are
self-conjugate and selection rules are available.

\end{abstract}

\newpage

\subsection*{1. Introduction}
An investigation of the quantum mechanical particles  in  the
external Dirac monopole's field has been carried  out  by  many  authors
(see, for example, in  [1-6]). Particularly, a~special  interest
was given to properties  of  these  systems  with  respect  to
the~operation of spatial $P$-inversion  [7-12]. As known, in virtue of
the~monopole-based
$P$-violation, the~usual particle's $P$-inversion operator
$\hat{\Pi}_{particle} \otimes \hat{P}$
does not commute with the~Hamiltonian $\hat{H}^{eg}$. The way of how to obtain
a~certain formal covariance of the~monopole-containing system  with
respect to $P$-symmetry there has been a~subject of special interest in
the~literature.
For instance: (a) those possibilities were discussed ([9]) in the
context of the generalized (allowing for the monopole presence)
$CPT$ theorem ($CPT \rightarrow CMPT \equiv  CNT$); (b)  in a number
of works (for example, see references [13-17]  it was  claimed
that this operator  plays a~role in  hierarchy  of  the~established
selection rules with respect to the~relevant  generalized  quantum
number $j$.

All the suggestions represent, in the~essence, a~single one:
the~magnetic charge is to be considered as a~pseudo scalar
quantity\footnote{One should take into account that this, as it is,
applies only to the~Schwinger basis; the~use of the~Dirac gauge or any other,
except Wu-Yang's, implies quite definite modifications  in representation of
the~$P$-operation on the monopole 4-potential.}.
 For the~subject under consideration, this assumption implies that
one ought to accompany the~ordinary $P$-transformation with a~formal
operator  $\hat{\pi}$ changing the~parametre $g$ into $-g$.
Correspondingly, the~composite discrete operator
$
\hat{N} = \hat{\pi} \otimes \hat{\Pi}_{particle} \otimes \hat{P}
$
will commute with the~relevant Hamiltonian.

Analysis of certain aspects of that monopole $P$-asymmetry
constitutes a~basic  goal  of the~present paper. Technical innovation
of the exposition below
is the~use of wave equations in the frame of the tetrad
formalism of Tetrode-Weyl-Fock-Ivanenko [18,19]. At this,
the Dirac ($S = 1/2$) and
Duffin-Kemmer ($S = 1$) equations  are
referred to a~basis of diagonal spherical tetrad; correspondingly, we
will use explicit forms of wave functions referring to the same
 tetrad basis (a~monopole potential is taken in Schwinger's form; we
 adhere designations used in [20,21]).

In Sec.2, several
facts on properties of spin $S = 1/2$  particle's wave functions
affected by external  monopole  field  are  briefly remembered.
Particularly, it is noted that there exists a~discrete operator
replacing the ordinary $P$-reflection:  $ \hat{N}_{bisp.} = \hat{\pi
} \otimes  \hat{\Pi}_{bisp.} \otimes \hat{P}$ which  commutes  with
the  Hamiltonian  and  can  be diagonalized on the~wave functions
$\Psi ^{eg.S=1/2}_{\epsilon jm}(x)$.  In Sec.3  the case of $S = 1$
is considered; here also there is an~operator $\hat{N}_{vect.}$:
$\hat{N}_{vect.} = \hat{\pi } \otimes \hat{\Pi}_{vect.} \otimes
\hat{P}$; but, in contrast
to  the $S = 1/2$ case,   the~$\hat{N}_{vect.}$   cannot   be
diagonalized   on  the~functions $\Psi ^{eg.S=1}_{\epsilon jm}(x)$.
So,  these two systems  exhibit  sharp distinction.
In Sec.4, two questions
are analyzed.  The~first one is the~property of non-self-conjugacy for
the~discrete  operators constructed for those $eg$-systems.
The~second   is  the~non-existence  of  any
$N$-parity selection rules, though the $\hat{N}_{bisp.}$ can be
diadonalized  on the relevant wave functions. As evidenced in Sec.4,
this operator $\hat{N}_{bisp.}$  does not result in a basic
structural condition
$$
 \Psi (t, -\vec{r}) = (4 \times 4 - matrix) \;
\Psi (t, \vec{r})
\eqno(1a)
$$

 \noindent which would guarantee indeed the~existence of
certain selection rules with respect to the~discrete quantum
number. Instead, there
arises only the~following one:
$$
\Psi^{+eg}(t, -\vec{r}) =
(4 \times 4 - matrix) \; \Psi^{-eg}(t, \vec{r})
\eqno(1b)
$$

\noindent take notice of a~change
at $eg$ parametre:  this minor alteration is completely detrimental
to the~possibility of producing any selection rules.
Else one added fact is emphasized: the~radial system of equations at
$S = 1/2$ case only depends on the~modulus of the~parametre $eg$,
whereas in the $S = 1$ case it depends on the~sign of the~$eg$ too.
Evidently, it may be thought as indication  that the~formal
diagonalizing of the $\hat{N}_{bisp.}$ (and non-diagonalizing of
$\hat{N}_{vect.}$) correlates just with the latter circumstance.

Sec.5 treats briefly some facts on discrete symmetry in the
non-Abelian model: an isotopic doublet of Dirac fermions is discussed.
Here, the relevant discrete operator (containing $P$-inversion) is
self-conjugated, and correspondingly selection rules on a composite
(isotopic-Lorentzian) parity are available.  In Sec.6, the case of
isotopic doublet of vector particles in the external t'Hooft-Polyakov
potential is considered. The account is given of how the discrete
operator simplifies corresponding wave functions and how the system
of radial equations fits well  with limitations  imposed
on the functions by diagonalization of this operator.  It may be
noticed that just those mathematical relations which have supplied
``bad'' peculiarities in the Abelian theory have produced, in another
background, ``good'' characteristics of the
corresponding non-Abelian problems. So, the paper reveal the
interplay between Abelian and non-Abelian models regarding their
properties under discrete symmetry.  Sec.7 provides some more
discussion on possible implications of monopole-based
$P$-(a)symmetry.

\subsection*{2. $eg$-system at $S = 1/2$}

The generally relativistic Dirac equation
in the chosen basis has the form [20]
$$
\left [\; i \gamma ^{0} \; \partial _{t} \; + \; i \gamma ^{3}\;
 (\;\partial _{r} \; + \; {1 \over r} \;) \; + \;
{1\over r} \; \Sigma ^{\lambda}_{\theta ,\phi } \; - \; {mc \over \hbar}\;
\right ] \; \Psi (x) = 0
\eqno(2a)
$$

\noindent where
$$
\Sigma ^{\lambda}_{\theta ,\phi } =
i \gamma ^{1}\; \partial _{\theta } \; + \; \gamma ^{2} \;
{{i\partial _{\phi }\; + \; ( ij^{12} - \lambda )\; \cos \theta}
\over{ \sin \theta}}
\eqno(2b)
$$

\noindent and $\lambda  = eg/ \hbar c $. The~wave function with quantum
numbers  $\epsilon , \; j, \; m $ (See all details in [20]) is
$$
\Psi _{\epsilon jm}(t,r,\theta ,\phi ) =
{{e^{-i \epsilon t} } \over r }
\left ( \begin{array}{l}
                        f_{1} \; D_{\lambda - 1/2} \\
                        f_{2} \; D_{\lambda + 1/2} \\
                        f_{3} \; D_{\lambda - 1/2} \\
                        f_{4} \; D_{\lambda + 1/2}
                                                \end{array} \right )
\eqno(3)
$$

\noindent the symbol $D_{\sigma}$ denotes the Wigner functions:
$D_{\sigma} \equiv  D^j _{-m, \sigma}( \phi, \theta, 0)$.
For $\lambda $ and $j$, only the~following values are allowed:
$$
 \lambda  = eg /\hbar c  = \;\pm  1/2, \; \pm  1, \;\pm  3/2, \ldots
\qquad and \qquad
j = \; \mid \lambda \mid -1/2,  \; \mid \lambda \mid + 1/2 , \;
     \mid \lambda \mid +3/2, \ldots
\eqno(4)
$$

\noindent correspondingly the substitution (3) is valid  only for
$j > j_{min.} = \mid \lambda \mid -1/2$.
The case of minimal allowable value $j_{min.}= \mid k \mid - 1/2$  must  be
separated out and looked into in a special way.  For example, let
$\lambda = \pm 1/2 $, then to the minimal value $j = 0$ there
correspond the~wave functions $$ \Psi ^{(j=0)}_{\lambda = +1/2}(x) =
{ e^{-i\epsilon t} \over  r} \left ( \begin{array}{l} f_{1}(r)  \\
           0 \\  f_{3}(r)  \\  0  .  \end{array} \right ) \;\; , \;\;
\Psi ^{(j =0)}_{\lambda = -1/2}(x) =
{e^{-i\epsilon t} \over  r}
\left ( \begin{array}{l}
   0  \\  f_{2}(r)  \\   0   \\  f_{4}(r)
\end{array} \right )        \;       .
\eqno(5)
$$

\noindent Thus, if $\lambda = \pm  1/2$, then to the minimal allowed
values $J_{\min }$ there correspond the function substitutions which
do not depend at all on the angular variables $(\theta ,\phi )$; at
this point there exists some formal analogy between  these
electron-monopole  states  and $S$-states ( with $l = 0 $) for
a~boson field of spin zero:  $\Phi _{l=0} = \Phi (r,t)$. However, it
would be unwise to attach too much significance to this formal
coincidence  because such $(\theta ,\phi )$- independence of
$(e-g)$-states  is  not  a~fact  invariant   under   tetrad   gauge
transformations. In contrast, the relation
$
\Sigma^{\pm 1/2}_{\theta ,\phi } \; \Psi ^{(j=0)}_{\lambda =\pm 1/2}
(x) \; \equiv  0 $ is gauge invariant.   Correspondingly,
the~matter equation above takes on the form
$$
 [\;i\; \gamma ^{0} \;
\partial_{t} \; + \; i\; \gamma ^{3} \; (\partial_{r} \; + \;  {1
\over r}\;  )\; - \;  mc/\hbar \; ] \; \Psi ^{(j=0)} = 0 \; .
\eqno(6)
$$

\noindent It is readily  verified  that  both functions in (5)
are directly extended to $(e-g)$-states  with $j = j_{\min }$ at all
the other $\lambda =\pm 1, \pm 3/2, \ldots $.
Indeed,
 $$
\Psi ^{\lambda > 0} _{j_{min.}} (x) =
{e^{-i\epsilon t} \over r} \left ( \begin{array}{l}
f_{1}(r) \; D_{\lambda-1/2}  \\  0  \\
f_{3}(r) \;  D_{\lambda-1/2} \\  0
\end{array} \right ) \; \; , \;\;
\Psi ^{\lambda <0} _{j_{min.}} (x) = { e^{-i\epsilon t} \over  r}
\left ( \begin{array}{l}
    0    \\   f_{2}(r) \; D_{\lambda +1/2}  \\  0  \\
    f_{4}(r) \; D_{\lambda +1/2}    \; ;
\end{array} \right )
\eqno(7)
$$

\noindent and,  as  can be shown,  the relation
$\Sigma^{\lambda} _{\theta ,\phi } \Psi _{j_{\min }}\equiv 0 $ still
holds.

After separating the variables, the radial system
is  ($ \nu  =  \sqrt {(j + 1/2)^{2} - \lambda ^{2}} $; for simplicity,
here let  us restrict ourselves to the~non-minimal $j$ states)
$$
\epsilon
\; f_{3} \; - \; i \; {d\over dr} \;f_{3} \; - \;
 i\;{\nu \over r}\; f_{4} \; - \; m \; f_{1} = 0 \; , \qquad
\epsilon  \; f_{4} \; + \; i \; {d\over dr} \;f_{4} \; + \;
i\;{\nu \over r} \; f_{3} \; -\; m \; f_{2} = 0  \; ,
$$ $$
\epsilon
\; f_{1} \; + \; i\; {d\over dr} \;f_{1} \; + \; i \; {\nu \over r}\;
f_{2} \; - \; m \; f_{3} = 0 \; ,  \qquad
\epsilon\; f_{2}\; - i\; {d\over dr} \;f_{2} \; - \; i\;
{\nu \over r}\; f_{1} \; - \; m \;f_{4} = 0 \; .
\eqno(8)
$$

\noindent As can be readily shown, on the functions (3) it is possible to
diagonalize a~discrete operator constructed on the~base of the~usual
bispinor $P$-reflection.  This  $P$-reflection
in  the  Cartesian tetrad basis is
$$
\hat{P}^{Cart.} = \hat{\Pi}^{Cart}_{bisp.} \otimes \hat{P} \; , \;\;
\hat{\Pi}^{Cart.}_{bisp.} =
\left ( \begin{array}{cccc} 0 &  0 &  i &   0  \\ 0 &
0 &  0 &   i  \\ i & 0 &  0 &   0  \\ 0 &  i &  0 &   0 \end{array}
\right )   \;  , \;\;
\hat{P} (\theta , \phi ) = (\pi  - \theta, \; \phi+ \pi ) \; ,
$$

\noindent being subjected to translation into the spherical one,
$
\hat{ P}^{sph.} = S(\theta ,\phi ) \;\hat{P}^{Cart.}\;S^{-1}(\theta
,\phi )$ gives us
$$
\hat{ P}^{sph.} \; \; = \hat{\Pi}^{sph.}_{bisp.} \otimes \hat{P} \; ,
\;\; \hat{\Pi}^{sph.}_{bisp.} = \left (\begin{array}{cccc} 0 &  0 &
0 & -1   \\ 0 &  0 & -1 &  0   \\ 0 & -1&  0 &  0   \\ -1&  0 &  0 &
0 \end{array} \right ) \;  .  $$

\noindent A required operator is of the form
$
\hat{N}_{bisp.} = \hat{\pi } \otimes \hat{\Pi}^{sph.}_{bisp.} \otimes
\hat{P}; $ here, $\hat{\pi }$ is a~special formal operation changing
$+eg$ into $-eg$,  and conversely:  $\hat{\pi } F( \lambda  ) = F( -
 \lambda )$.  From the~equation on proper values $\hat{N}_{bisp.} \;
 \Psi ^{\lambda }_{\epsilon jm}   = N \;  \Psi ^{\lambda }_{\epsilon
 jm}$   it follows   ($  \delta = \pm 1  $)
$$
N = \delta(-1)^{j+1}    \;: \qquad f_{4} = \delta \; f_{1} \; ,  \qquad f_{3} = \delta \;
f_{2} \eqno(9)
$$

\noindent these limitations are compatible with the~radial system (8).
It  should be emphasized that
some unexpected peculiarities with that procedure, in reality,  occur
as we  turn  to the states of minimal values of $j$.
Actually, let $\lambda = + 1/2 $ or $-1/2$  ($j  = 0$); then from the
equation    on proper values $\hat{N} \; \Psi ^{(j=0)} =
N \; \Psi ^{(j=0)}$  it follows
$$
\left ( \begin{array}{r} 0  \\  -
f_{3} \\  0 \\ - f_{4} \end{array} \right ) \; = \; N \; \left (
\begin{array}{r} f_{1}\\  0 \\ f_{3} \\  0 \end{array} \right ) \; ,
\;\; or  \;\;
\left ( \begin{array}{l} -f_{4} \\  0  \\ -f_{2} \\  0
\end{array} \right ) = \; N \; \left ( \begin{array}{r} 0 \\ f_{2} \\
0  \\ f_{4} \end{array}  \right )   \; .
$$

\noindent Evidently, they both  have no solutions,
 excluding trivially null ones (and therefore being of~no interest). Moreover,
as  may  be easily seen, in both cases
the~function $\Phi (x)$,  defined by
$\hat{N} \; \Psi ^{(j =0)}  \equiv  \Phi (x)$,
lies outside a~fixed totality of states that are only valid as allowed quantum
states of the system under consideration.
At greater values of this $\lambda$, we come  to  analogous
relations.

It should be useful to notice that the~above  simplification
$( \Psi _{\epsilon jm} \rightarrow  \Psi _{\epsilon jm\delta } )$  can
also  be   obtained   through   the~diagonalization of the~so-called
generalized Dirac operator  $\hat{K}^{\lambda}$
$$
\hat{K}^{\lambda}\; = \;  - \gamma ^{0} \gamma ^{3} \;
 \Sigma ^{\lambda}_{\theta ,\phi }  \;   .
\eqno(10a)
$$

\noindent Actually, from  $\hat{K}^{\lambda}
 \; \Psi _{\epsilon jm} (x) =
K \; \Psi _{\epsilon jm}$ we produce ($\delta \; = \pm  1$)
$$
K = - \delta \; (j+1/2) \; : \qquad
f_{4} = \; \delta \; f_{1} \; , \qquad f_{3} =\;\delta \; f_{2} \; .
\eqno(10b)
$$

\noindent
In turn, as regards the operator $\hat{K}^{\lambda}$ for the $j_{\min
 }$ states  we get $\hat{K}^{\lambda} \; \Psi _{j_{min.}} =  0$ ;
that is, this state represents the  proper  function  of  the $\hat{K}$
with the null  proper  value.  So,  application  of  this $\hat{K}$
instead of  the $\hat{N}$   has  an  advantage  of  avoiding  the
paradoxical and puzzling situation when $\hat{N} \;
\Psi ^{(j_{min})} \not\in  \{ \Psi  \}$.
In a sense, this second alternative ( the use  of $\hat{K}^{\lambda}$
 instead
of $\hat{N}$ at separating the variables and  constructing  the  complete
set of mutually commuting operators) gives us a possibility not to
attach great significance to the monopole discrete operator $\hat{N}$  but
to focus our  attention  solely  on  the  operator
$\hat{K}^{\lambda}$.

\subsection*{3. $eg$-system at $S = 1$}

The basic Duffin-Kemmer equation is [21]
$$
\left [\;  i \beta ^{0} \; \partial _{t} \; + \; i \; (\; \beta ^{3}\;
 \partial _{r} \; + \;
{1\over r} \; (\; \beta ^{1} \; j^{31} \; + \; \beta ^{2}\; j^{32})\; )\; + \;
{1\over r}\; \Sigma ^{\lambda }_{\theta ,\phi } \; -
\; {mc \over \hbar}\; \right ] \; \Phi (x) = 0 \; ;
\eqno(11)
$$
$$
\Sigma ^{\lambda }_{\theta ,\phi } =
\left [ \;i \beta ^{1} \; \partial _{\theta } \; + \;
  \beta ^{2} \; {{ i\partial _{\phi } \; + \;
(\; ij^{12} \; - \; \lambda\; ) \; \cos\theta} \over
{ \sin \theta} } \; \right ] \; .
\eqno(12)
$$

\noindent The wave functions with quantum numbers $(\epsilon, j, m)$
can be taken in the form
$$
\Phi _{\epsilon jm}(x)= e^{-i \epsilon t} \;
\left [\; f_{1}(r) \; D_{\lambda } \; , \;  f_{2}(r) \; D_{\lambda -1} \; ,\;
 f_{3}(r) \; D_{\lambda } \; ,\; f_{4}(r) \; D_{\lambda +1} \; , \;
f_{5}(r) \; D_{\lambda -1} \right. \; ,
$$
$$
 \left. f_{6}(r)  \; D_{\lambda } \; , \; f_{7}(r) \; D_{\lambda +1} \; ,  \;
  f_{8}(r) \; D_{\lambda -1}\; , \;  f_{9}(r) \; D_{\lambda }\; , \;
  f_{10}(r) \; D_{\lambda +1} \; \right ] \;
\eqno(13)
$$

\noindent here, as above, $D_{\sigma} = D^{j}_{-m, \sigma }(\phi ,\theta ,0)$.
For  quantities $\lambda$  and $j$, the values are allowed
$$
1.\qquad  if \qquad \lambda = \pm 1/2 \; ,  \qquad
then \qquad j = \; \mid \lambda \mid \; , \; \mid \lambda \mid  + 1 \; , \ldots ;
$$
$$
2.\qquad  if \qquad \lambda = \; \pm 1 \; , \; \pm 3/2, \ldots \; , \qquad
then \qquad j = \; \mid \lambda \mid  - 1 \; , \; \mid \lambda \mid \; ,
\; \mid \lambda \mid +1 \; , \ldots
$$

\noindent Correspondingly, the substitution (13) is applied  only  to
the non-minimal $j$ values; for simplicity, let us consider just  those
states. After separation of variables we get
$$
-(\; {d \over {dr}} \; + \; {2\over r}\; )\;  f_{6} \; - \;
\sqrt{2} \; {1\over r} \; (\; c \; f_{5} \; + \; d \; f_{7}\; )\;
 \; - \; m \; f_{1} = 0  \; ,
$$$$
i \epsilon \; f_{5} \; + \; i (\; {d \over{dr}} \; +
\; {1\over r}\; ) \; f_{8} \; + \;
i \sqrt{2} \; {c\over r} \; f_{9} \; - \; m \; f_{2} = 0  \;   ,
$$
$$
i \epsilon \; f_{6} \; + \; {2i\over r}\; (\; -c \; f_{8} \; +
\; d \; f_{10}\;) \; -\;  m f_{3} = 0  \; ,
$$
$$
i\epsilon \; f_{7} \;  - \; i (\; {d \over{dr}} \; + \; {1 \over r}\; )\;
 f_{10} \;  -  \; i \sqrt{2} \; {d \over r} \; f_{9} \; -
\; m \; f_{4} = 0 \; ,
$$
$$
i \epsilon \; f_{2} \; + \; \sqrt{2} \; {c \over r}\;  f_{1} \;
- \; m \; f_{5} = 0 \; ,
$$
$$
-i \epsilon \; f_{3} \; - \; {d \over{dr}} \; f_{1} \; -  \; m \; f_{6} = 0 ,
$$
$$
i \epsilon \; f_{4} \; + \; \sqrt{2} \; {d  \over r} \; f_{1} \; - \;
m \; f_{7} = 0 \; ,
$$
$$
-i\; (\; {d\over {dr}}\; + \; {1\over
r}\; )\; f_{2} \; - \; i \sqrt{2}\; {c \over r} \; f_{3} \; - \; m \;
f_{8} = 0  \; ,
$$
$$
 i \sqrt{2}\; {1\over r} \; (\; c \; f_{2} \; -
\; d \; f_{4}\; ) \; - \; m \;f_{9} = 0  \;,
$$
$$
i\; (\; {d \over
{dr}\; } \; + \; {1\over r}\; )\; f_{4} \; + \; i\sqrt{2}\; {d \over
r}\; f_{3} \; - \; m \; f_{10} = 0
\eqno(14)
$$

\noindent where $
 c={1\over 2} \sqrt{(j+ \lambda)( j - \lambda +1)} , \qquad
 d={1\over 2} \sqrt{(j- \lambda)( j + \lambda  +1)}$.

As in case of  a  fermion  field  above, here we try to
use a~generalized operator $\hat{N}_{vect.}$  commuting  with
the wave operator in (11).
The  vector  ordinary $P$-reflection in  Cartesian
tetrad, is
$$
\hat{P}^{Cart.} \; = \; \hat{\Pi}^{Cart.}_{vect.} \otimes \hat{P} \;
, \;\; \hat{\Pi}^{Cart.}_{vect.} = \left ( \begin{array}{cccc} 1  &
     0  &  0  &  0  \\ 0  &  -I  &  0  &  0  \\ 0  &   0  & -I  &  0
     \\ 0  &   0  &  0  & +I \end{array} \right )
\eqno(15a) $$

\noindent where  a symbol "I" denotes a~unit $3\times 3$ matrix.
After translating this $\hat{P}^{Cart.}$ into the~spherical
tetrad's  basis  according  to
$ \hat{P}^{sph.} = O(\theta ,\phi ) \;
  \hat{P}^{Cart.}\;
 O^{-1}(\theta ,\phi )$,
where $( O(\theta ,\phi)$  is  a  $10$-dimension
rotational  matrix associated with taking  the~Cartesian gauge into
the spherical one), it takes on the~form (the standard cyclic basis
in the vector space is used)
$$
\hat{P}^{sph.} \; = \; \hat{\Pi}^{sph.}_{vect.} \otimes \hat{P} \; , \;\;
\hat{\Pi}^{sph.}_{vect.}  =
\left ( \begin{array}{cccc}
     1  &   0  &  0  &  0  \\
     0  &  +E  &  0  &  0  \\
     0  &   0  & +E  &  0  \\
     0  &   0  &  0  & -E
\end{array} \right ) \; , \;\;
E \equiv \left ( \begin{array}{ccc}
          0   &  0  & 1 \\ 0  & 1  &  0 \\  1  &  0  &  0
\end{array} \right )
\eqno(15b)
$$

\noindent A required operator (in the spherical basis) is
$$
\hat{N}^{sph.}= \hat{\pi } \otimes \hat{\Pi}_{vect.}^{sph.}
\otimes \hat{P}
\eqno(15c) $$

\noindent From the equation
$\hat{N}_{vect.} \; \Psi ^{\lambda}_{\epsilon jm} =
 N  \; \Psi ^{\lambda}_{\epsilon jm}$
we get
$$
N = (-1)^{j+1}\; : \qquad  f_{1} = f_{3} = f_{6} = 0 \; ,\;\;\;  f_{4}= - f_{2}
\; ,\;\;\; f_{7} = - f_{5} \; , \;\;\; f_{10} = + f_{8} ;
\eqno(16a)
$$
$$
N = (-1)^{j} \; :  \qquad f_{9}= 0 \qquad f_{4} = + f_{2}, \qquad
f_{7} = + f_{5} , \qquad  f_{10} = - f_{8} \; .
\eqno(16b)
$$

In contrast to the~fermion case above,  here   the~relations
(16a,b) are readily shown not to be  compatible  with  the~radial
system (14). However, as can be  easily  verified,  this  operator
indeed commutes with the~wave operator in  (11).  Thus,  apparently
there exists  a~ contradiction. So different  properties   of
particles with spin 1/2 and 1 in external  monopole  field, while
one notes their complete origin  similarity,
seem to be rather surprising and puzzled.

\subsection*{4. $N$-operator and property of self-conjugacy}

So, in both cases $S  =  1/2$  and $S  = 1$ , the~respective
$N$-operators are constructed in accordance with the same pattern:
$$
\hat{N}= [\;  \hat{\pi } \otimes \hat{\Pi}_{particle}
 \otimes \hat{P} \;] \; , \;\;
[ \hat{N} , \; \hat{H} ^{eg} ]_{-}= 0
\eqno(17) $$

\noindent where  $\hat{\Pi}_{particle} = \hat{\Pi}_{bisp.}$ or
$\hat{\Pi}_{vect.}$, and $\hat{H}^{eg} = \hat{H}^{eg}_{bisp.}$  or
 $\hat{H}^{eg}_{vect.}$, respectively. However, as was just noted,
there are some essential distinctions between these two situations,
and this deserves special consideration.
At  a~glance, the~situation at $S = 1$ looks as very
contrasting with all generally accepted concepts of
the~conventional quantum mechanics. Indeed, the~commutation  relation
required $[ \; \hat{N}_{vect.} , \hat{H}^{eg}_{vect.} \; ]_{-} = 0$  holds,
but this $\hat{N}_{vect.}$   is not diagonalized onto
$H^{eg}_{vect.}$'s  eigenfunctions
$\Phi ^{eg}_{ejm}$. As regards
to $S = 1/2$ situation, that (as would be seemed) entirely comes
under the common and familiar requirements of quantum theory.
However, on more closing  consideration,  it  will  be  clear
that,  first, $S = 1$   situation   does   not   turn   out   to
contradict  the  commonly  acknowledged  requirements  of  quantum
mechanics; second, the $S = 1/2$ situation does not provide us just
else one trivial illustration to the familiar interrelation of the
commutation rule $[ \hat{A} , \hat{H} ]_{-} = 0$ and
the~possibility to measure
simultaneously those quantities  $\hat{A}$ and $\hat{H}$.

All above, as a~correcting and revealing remark, it  must  be
stressed that  the  quantum  mechanics,  when  dealing  with  some
specific operator  $\hat{A}$, implies  essentially  its  self-conjugacy
property:
$
< \Psi  \mid  \hat{A}\;\Phi >\;=\;< \hat{A}\; \Psi \mid \Phi >.
$
For  example,  the~usual  bispinor $P$-reflection
presents evidently a~self-conjugate one,  since  one  has
$$
<\Psi (\vec{r}) \mid  \gamma ^{0} \; \hat{P} \; \Phi (\vec{r})> \; =
\int \tilde {\Psi } ^{*}(\vec{r}) \; \Phi (-\vec{r}) \;dV  \; ,
$$$$
< \gamma ^{0} \; \hat{P} \; \Psi (\vec{r}) \mid \Phi (\vec{r})>\; =
\int \tilde{\Psi }^{*}(- \vec{r}) \; \Phi (\vec{r}) \; dV   \; .
$$

\noindent
The~$\Psi $ with over symbol $\sim$  denotes a~transposed  column-function,
that is, a~row-function; and   the~asterisk  $*$   designates  the
operation of complex conjugation.
In the presence of the external monopole field,  the~whole
situation is completely different from the above, namely,  the $\hat{N}$
used here does not possess the required self-conjugacy property. Indeed,
$$
<\psi ^{+eg}(\vec{r}) \mid  \hat{N} \; \Phi ^{+eg}(\vec{r}) >\; =
\int (\tilde {\Psi }^{+eg}(\vec{r}))^{*} \; \Phi ^{-eg}(-\vec{r}) dV
\; ,
$$
$$
< \hat{N} \; \Psi ^{+eg}(\vec{r}) \mid \;\Phi ^{+eg}(\vec{r})>\; =
\int (\tilde  {\Psi }^{-eg}(\vec{r}))^{*} \; \Phi ^{+eg}(-\vec{r}) \;
dV \; .  $$

\noindent It is evident at a~glance  that  right  hand  sides  of  these  two
equalities vary in sign at $eg$ parametre; thereby it follows  that
the~discrete operator $\hat{N}$  does not possess
the~self-conjugacy property. As regards to such a~property  of $\hat{N}$,
the case of $S = 1$ looks  completely  alike.  This  peculiarity  of
$\hat{N}_{bisp.}$  and $\hat{N}_{vect.}$ may be interpreted as
follows:  those $\hat{N}$ do not afford any physical observables
which could be measured  by any physical apparatus.  In other words,
the features of $S = 1$ case  mentioned  above do not go into
contradiction with  proper  principles  of  the quantum theory.  On
the other hands, one  could  acknowledge  oneself  puzzled when only
specializing to $S = 1/2$ system. In the latter case,  as it would
seems, the familiar connection between commutation relations and
measuring the $\hat{N}$ is realized. But such a natural  reference
to this  familiar arrangement is not valid  here because  of  already
mentioned arguments of non-self-conjugacy; and what is more, the
existence of contrasting situations at $S = 1/2$ and $S = 1$  directly
suggests that one must attach more significance  to  the  latter
(of non-self-conjugacy) requirement.
In this connection, one must take notice  of  the  manner  in
which  the $eg$   parametre  enters   the   radial    system   for
$f_{1},\ldots, f_{4}$ : it  occurs through
 $\nu  = \sqrt {(j+1/2)^2 - \lambda^2 }$.
The latter leads to independence on $\lambda's$   sign.  Therefore,
the two distinct systems   with  the  characteristics $+eg$   and
$-eg$ respectively have their radial systems exactly identical:  $$
F^{+eg}_{s=1/2}(f_{1},\ldots, f_{4}) =
F^{-eg}_{s=1/2}(f_{1},\ldots, f_{4}) \; .
\eqno(18)
$$

\noindent In contrast to this, the $S = 1$ affords an  essentially  different
case: here, the parametre $eg$  enters the relevant  radial  system
through $c$ and $d$ , that is,
two  radial systems marked by $+eg$  and $-eg$  respectively,
though can easily be inverted into each  other  by  simple  formal
procedure, vary in their explicit form:
$$
F^{+eg}_{s=1} (f_{1},\ldots, f_{10}) \neq
F^{-eg}_{s=1} (f_{1},\ldots, f_{10})  \; .
\eqno(19)
$$

As an~illustration to manifestations of the non-self-conjugacy property
of  the $N$-operator,  let  us  consider a~question  concerning $P$-parity
selection rules  in  presence  of  the~monopole.   Though  at
this situation  there  exists  an~operator  commuting  with
 the~Hamiltonian:
$$
\hat{N} = \hat{\pi } \otimes  \hat{\Pi}_{bisp.} \otimes
\hat{P} \; , \qquad \hat{N} \;  \Psi ^{eg}_{\epsilon jm\mu }(\vec{r}) =
\mu \; (-1)^{j+1} \; \Psi ^{eg}_{\epsilon jm\mu }(\vec{r})
\eqno(20)
$$

\noindent $( \hat{\pi } \; \Psi ^{+eg}_{\epsilon jm\mu }(\vec{r}) =
\Psi ^{-eg}_{\epsilon jm\mu }(\vec{r})\; )$,
but this does not allow us to  obtain
any $N$-parity selection rules. Let us  consider  this  question  in
more detail.
A matrix element for some physical observable $\hat{G}^{0}(x)$   is  to
be
$$
\int \bar{\Psi}^{eg}_{\epsilon jm\mu }(\vec{r}) \;
\hat{G}^{0}(\vec{r})\;\Psi ^{eg}_{\epsilon j'm'\mu'}(\vec{r})\; dV
\equiv \int r^{2}dr \int f( \vec{r} ) \; d\Omega  \; .
\eqno(21)
$$

\noindent
First we  examine the case $eg = 0,$ in order  to  compare  it
with the situation at $eg \neq  0$. Let us relate $f(-\vec{r})$
with $f(\vec{r})$. Considering the equality  (and the same  with
$j'm' \delta ')$
$$
\Psi ^{0}_{\epsilon jm\delta }(-\vec{r} ) =
\hat{\Pi}_{bisp.} \; \delta \; (-1)^{j+1} \;
  \Psi ^{0}_{\epsilon jm\delta }(\vec{r} )
\eqno(22a)
$$

\noindent we  get
$$
f^{0}(-\vec{r}) = \delta  \; \delta '\; (-1)^{j+j'+1}
\bar{\Psi}^{0}_{\epsilon jm\delta }(\vec{r})  \;
 \left [\; \hat{\Pi}_{bisp.} \; \hat{G}^{0}(-\vec{r}) \;
  \hat{\Pi}_{bisp.} \; \right ]
\; \Psi ^{0}_{\epsilon j'm'\delta '}(\vec{r}) \; .
$$

\noindent If $\hat{G}^{0}(\vec{r})$  obeys  the equation
$$
[\; \hat{\Pi}_{bisp.} \; \hat{G}^{0}(-\vec{r}) \;
 \hat{\Pi}_{bisp.} \; ] =
 \omega ^{0} \; \hat{G}^{0}(\vec{r} )
\eqno(22b)
$$

\noindent here $\omega ^{0}$ defined to be $+1$   or   $-1$  relates  to
the  scalar  and pseudo scalar, respectively, then $f(\vec{r})$
can be brought to
$
f^{0}(-\vec{r}) = \omega \;  \delta \; \delta '\; (-1)^{j+j'+1} \;
f^{0}(\vec{r}).
$
The latter can generate the well\-known $P$-parity selection rules:
$$
\int \bar{\Psi}^{0}_{\epsilon jm\mu }(\vec{r}) \; \hat{G}^{0}(\vec{r}) \;
\Psi ^{0}_{\epsilon j'm'\mu'}(\vec{r}) \;dV
= \left [\; 1 \;+ \; \omega \;  \delta \;  \delta '\; (-1)^{j+j'+1} \; \right ]
\; \int r^2 \;  dr \; \int_{1/2} f^{0}(\vec{r}) \;  d\Omega
\eqno(22c)
$$

\noindent where the $\theta,\phi$-integration is performed on a~half-sphere.
In contrast to everything just said, the~situation at $eg \neq  0$  is
completely different since any  equality in the  form $(22a)$ does
not appear here.  In other words, in virtue of the absence any
correlation between $f^{eg}(\vec{r})$ and $f^{eg}(-\vec{r})$,
there is no selection rules on discrete quantum number $N$.
In  accordance  with  this,
for instance, an expectation value for the usual operator of space
coordinates $\vec{x}$  need not equal zero and one follows this
(see in [13-17]).

In the same time, from the above  it follows that there exist quite
definite correlations between $\Psi^{\pm eg}(-\vec{r})$ and
$\Psi^{\mp eg}(\vec{r})$ as well as  between
$f^{\pm eg}(-\vec{r})$ and
$f^{\mp eg}(\vec{r})$ (supposedly, the relation (22b) still holds):
$$
\Psi^{\pm eg}(-\vec{r}) = \delta (-1)^{j+1} \;
\hat{\Pi}_{bisp.}\;
\Psi^{\mp eg}(\vec{r}) \;\; , \;\;
f^{\pm eg}(-\vec{r}) =
 \omega \;  \delta  \; \delta '\; (-1)^{j+j'+1} \;
f^{\mp eg}(\vec{r})  \; .
$$

\noindent Those latter provide  certain indications that
in a non-Abelian (monopole-contained) model no problems with discrete
$P$-inversion-like symmetry might occur.  In confirmation to this let
us consider some  facts on particle-monopole   systems in the
non-Abelian situation.

\subsection*{5. Doublet of fermions}

It can be shown that the wave functions for the doublet of Dirac particles in
the external monopole (t'Hooft-Polyakov's) potential can be constructed
in the form (for more detail see [22])
$$
\Psi_{\epsilon jm\delta} (t, r, \theta, \phi) = e^{-i \epsilon t} \;
\left [ T_{+1/2} \otimes
\left ( \begin{array}{c}
f_{1} \; D_{-1} \\ f_{2} \; D_{\;0} \\ f_{3} \; D_{-1} \\ f_{4} \; D_{\;0}
\end{array} \right ) \; + \; \delta \; T_{-1/2} \otimes
\left ( \begin{array}{c}
f_{4} \; D_{\;0} \\ f_{3} \; D_{+1} \\ f_{2} \; D_{\;0} \\ f_{1} \; D_{+1}
\end{array} \right ) \; \right ]
\eqno(23)
$$

\noindent they  represent  eigenfunctions of operators
$\vec{J}^{2}, J_{3}, \hat{N} =
\sigma_{1} \otimes \hat{\Pi}_{bisp.} \otimes \hat{P} \; ; \;
                                            $.
Here, the discrete
operator $\hat{N}$ provides a self-conjugated quantity. In addition,
the wave  functions obey the condition  ($ \delta  = \pm 1$)
$$
N = \delta (-1)^{j+1}\; : \; \;
\Psi_{\epsilon jm\delta}(-\vec{r}) =
\delta (-1)^{j+1} (\sigma_{1} \otimes \hat{\Pi}_{bisp.} )\;
\Psi_{\epsilon jm\delta}(+\vec{r})
$$

\noindent in virtue of that the corresponding selection rules are
available\footnote{These questions and a number of other ones
will be analyzed with much more details in a separate  paper of
the  author.}. In particular,  these
selection rules predict that the expectation value of the  spatial coordinates
will be equated to zero

$$
<\Psi_{\epsilon jm\delta} \mid \vec{r} \mid \Psi_{\epsilon jm\delta}>
\;  \sim [\; 1\; - \;\delta^{2} (-1)^{2j}\; ] \equiv 0     \; .
\eqno(24a)
$$

\noindent That vanishing can be readily understood from the following expansions
$$
\Psi_{\epsilon jm \delta } = [ \; T_{+1/2} \otimes \Psi^{+} \; + \;
                                T_{-1/2} \otimes \Psi^{-} \; ] \; ,
$$
$$
<\Psi_{\epsilon jm\delta} \mid \vec{r} \mid \Psi_{\epsilon jm\delta}> =
[\; <\Psi^{+} \mid \vec{r}\mid \Psi^{+}> \; + \;
<\Psi^{-} \mid \vec{r}\mid \Psi^{-}>\; ]
\eqno(24b)
$$

\noindent and the fitting relationship
$$
   \bar{\Psi}^{\pm}(-\vec{r})\; (-\vec{r}) \; \bar{\Psi}^{\pm}(-\vec{r}) =
-  \bar{\Psi}^{\mp}( \vec{r})\; ( \vec{r}) \; \bar{\Psi}^{\mp}( \vec{r}) \; .
\eqno(24c)
$$

\noindent That is, the ``bad'' mathematical relations (24c) in the
Abelian model turn out to be ``good''ones in the non-Abelian theory
background.

\subsection*{6. Doublet of vector particles}

Now, let us consider briefly  how  the problem of discrete symmetry looks
in the situation of the  vector particles
doublet in the t'Hooft-Polyakov
potential. Here, the matter equation is (the spherical tetrad basis and
the Shwinger unitary gauge in isotopic space are used)
$$
\left [\; i \beta ^{0} \; \partial _{t} \; + \; i (\; \beta ^{3}\;
 \partial _{r} \; + \; {1 \over r} \; (\beta^{1} J^{31} + \beta^{2} J^{32})
\; + \; {1\over r} \; \Sigma _{\theta ,\phi } \; + \;  \right.
$$
$$ \left. {e r^{2} K(r) + 1 \over r}\;
( t^{2} \otimes \beta^{1} - t^{1} \otimes \beta^{2}) \; -
\; {mc \over \hbar}\;
\right ] \; \Phi (x) = 0
\eqno(25a)
$$

\noindent where
$$
\Sigma _{\theta ,\phi } =
i \beta ^{1}\; \partial _{\theta } \; + \; \beta ^{2} \;
{{i\partial _{\phi }\; + \; ( ij^{12} + t^{3} )\; \cos \theta}
\over{ \sin \theta}} \;   \; .
\eqno(25b)
$$

\noindent The function $K(r)$ enters the non-Abelian monopole solution
$W^{(a)}_{i} = \epsilon_{iab} x^{b} K(r); t^{i} = {1 \over 2} \sigma^{i}$.
The composite wave function is to be [22]
$$
\Phi_{\epsilon ,jm} = e^{- i\epsilon t}\; \left [\;  T_{+1/2} \otimes
\left ( \begin{array}{c}
f_{1} \; D_{-1/2} \\
f_{2} \; D_{-3/2} \\   f_{3} \; D_{-1/2} \\   f_{4} \; D_{+1/2} \\
f_{5} \; D_{-3/2} \\    f_{6} \; D_{-1/2} \\   f_{7} \; D_{+1/2} \\
f_{8} \; D_{-3/2} \\    f_{9} \; D_{-1/2} \\   f_{10} \; D_{+1/2}
\end {array} \right )  \; + \; T_{-1/2} \otimes
\left ( \begin{array}{c}
g_{1} \; D_{+1/2} \\
g_{2} \; D_{-1/2} \\   g_{3} \; D_{+1/2} \\   g_{4} \; D_{+3/2} \\
g_{5} \; D_{-1/2} \\    g_{6} \; D_{+1/2} \\   g_{7} \; D_{+3/2} \\
g_{8} \; D_{-1/2} \\    g_{9} \; D_{+1/2} \\   g_{10} \; D_{+3/2}
\end {array} \right )  \;\right ]
\eqno(26)
$$

\noindent where $D_{\sigma} \equiv D^{J}_{-m,\sigma}(\phi, \theta,0)$;
the quantum   number $j$ takes values $1/2, 3/2, ...$ To separate
the variables,
actually new calculations (required in addition to the above Abelian
case) concern only the term proportional to
$[(e r^{2} K(r) + 1 )/ r ] \equiv W$ (just it mixes up two isotopic
components):
$$
( t^{2} \otimes \beta^{1} - t^{1} \otimes \beta^{2}) \;
\Phi_{\epsilon ,jm} = e^{- \epsilon t}\; \times
$$
$$
\left [ \;
T_{-1/2} \otimes
\left ( \begin{array}{c}
-\;  f_{7} \; D_{+1/2} \\
+i\; f_{9} \; D_{-1/2} \\   +i \; f_{10} \; D_{+1/2} \\   0  \\
+\; f_{1} \; D_{-1/2} \\    0  \\   0  \\
-i\; f_{3} \; D_{-1/2} \\   -i \; f_{4} \; D_{+1/2} \\   0
\end {array} \right )  \; + \; T_{+1/2} \otimes
\left ( \begin{array}{c}
-\; g_{5} \; D_{-1/2} \\
0 \\   -i \; g_{8} \; D_{-1/2} \\   -i \; g_{9} \; D_{+1/2} \\
0 \\    0  \\   +\; g_{1} \; D_{+1/2} \\
0 \\   +i g_{2} \; D_{-1/2} \\   +i g_{3} \; D_{+1/2}
\end {array} \right )  \; \right ] \; .
$$

After separation the variables, we  produce the equations on twenty
functions; we rearranged them in couple as convenient):
$$
-(\; {d \over {dr}} \; + \; {2\over r}\; )\;  f_{6} \; - \;
\sqrt{2} \; {1\over r} \; (\; c^{+} \; f_{5} \; + \; d^{+} \; f_{7}\; )\;
 \; - \; m \; f_{1} - \; W \; g_{5} = 0  \; ,
$$
$$
-(\; {d \over {dr}} \; + \; {2\over r}\; )\;  g_{6} \; - \;
\sqrt{2} \; {1\over r} \; (\; c^{-} \; g_{5} \; + \; d^{-} \; g_{7}\; )\;
 \; - \; m \; g_{1} - \; W \; f_{7} = 0  \; ;
$$
$$
i \epsilon \; f_{5} \; + \; i (\; {d \over{dr}} \; +
\; {1\over r}\; ) \; f_{8} \; + \;
i \sqrt{2} \; {c^{+} \over r} \; f_{9} \; - \; m \; f_{2} = 0  \;   ,
$$
$$
i \epsilon \; g_{7} \; - \; i (\; {d \over{dr}} \; -
\; {1\over r}\; ) \; f_{10} \; - \;
i \sqrt{2} \; {d^{-} \over r} \; g_{9} \; - \; m \; g_{4} = 0  \;   ;
$$
$$
i \epsilon \; f_{6} \; + \; {2i\over r}\; (\; -c^{+} \; f_{8} \; +
\; d^{+} \; f_{10}\;) \; -\;  m f_{3} \; - \; i W \; g_{8} = 0  \; ,
$$
$$
i \epsilon \; g_{6} \; + \; {2i\over r}\; (\; -c^{-} \; g_{8} \; +
\; d^{-} \; g_{10}\;) \; -\;  m g_{3} \; + \; i W \; f_{10} = 0  \; ;
$$
$$
i\epsilon \; f_{7} \;  - \; i (\; {d \over{dr}} \; + \; {1 \over r}\; )\;
 f_{10} \;  -  \; i \sqrt{2} \; {d^{+} \over r} \; f_{9} \; -
\; m \; f_{4} \;- \; i  W \; g_{9} = 0 \; ,
$$
$$
i\epsilon \; g_{5} \;  + \; i (\; {d \over{dr}} \; + \; {1 \over r}\; )\;
 g_{8} \;  +  \; i \sqrt{2} \; {c^{-} \over r} \; g_{9} \; -
\; m \; g_{2} \;+ \; i  W \; f_{9} = 0 \; ;
$$
$$
i \epsilon \; f_{2} \; + \; \sqrt{2} \; {c{+} \over r}\;  f_{1} \; -
\; m \; f_{5} = 0 \; ,
$$
$$
i \epsilon \; g_{4} \; + \; \sqrt{2} \; {d{-} \over r}\;  g_{1} \; -
\; m \; g_{7} = 0 \; ;
$$
$$
-i \epsilon \; f_{3} \; - \; {d \over{dr}} \; f_{1} \; -  \; m \; f_{6} = 0
\; ,
$$
$$
-i \epsilon \; g_{3} \; - \; {d \over{dr}} \; g_{1} \; -  \; m \; g_{6} = 0 \; ;
$$
$$
i \epsilon \; f_{4} \; + \; \sqrt{2} \; {d^{+} \over r} \; f_{1} \; -
\; m \; f_{7} \; + \; W \; g_{1} = 0  \; ,
$$
$$
i \epsilon \; g_{2} \; + \; \sqrt{2} \; {c^{-} \over r} \; g_{1} \; -
\; m \; g_{5} \; +\; W \; f_{1} = 0  \; ;
$$
$$
-i\; (\; {d\over {dr}}\; + \; {1\over r}\; )\; f_{2} \; - \;
i \sqrt{2}\; {c^{+} \over r} \; f_{3} \; - \; m \; f_{8} = 0  \; ,
$$
$$
+i\; (\; {d\over {dr}}\; + \; {1\over r}\; )\; g_{4} \; + \;
i \sqrt{2}\; {d^{-} \over r} \; g_{3} \; - \; m \; g_{10} = 0  \; ;
$$
$$
i \sqrt{2}\; {1\over r} \; (\; c^{+} \; f_{2} \; - \; d^{+} \; f_{4}\; ) \; -
 \; m \;f_{9} \; + \;i W \; g_{2} = 0  \; ,
$$
$$
i \sqrt{2}\; {1\over r} \; (\; c^{-} \; g_{2} \; - \; d^{-} \; g_{4}\; ) \; -
 \;  \;g_{9} \; + \; i W \; f_{4} = 0  \; ;
$$
$$
i\; (\; {d \over {dr}\; } \; + \; {1\over r}\; )\; f_{4} \; + \;
i\sqrt{2}\; {d^{+} \over r}\; f_{3} \; - \; m \; f_{10} \; +
\; i W \; g_{3} = 0    \; ,
$$
$$
-i\; (\; {d \over {dr}\; } \; + \; {1\over r}\; )\; g_{2} \; - \;
i\sqrt{2}\; {c^{-} \over r}\; g_{3} \; - \; m \; g_{8} \; -
\; i W \; f_{3} = 0 \; .
\eqno(27)
$$

\noindent where (see Sec. 3)
$c={1\over 2} \sqrt{(j+ \lambda)( j - \lambda +1)}\; , \;
d={1\over 2} \sqrt{(j- \lambda)( j + \lambda  +1)}$ and the signs $+$ (plus)
and $-$ (minus) relate to the $\lambda = -1/2$ and $\lambda = + 1/2$
respectively.
It is easily verified that the composite discrete operator
$
\hat{N} = (\; \sigma_{1} \otimes \hat{\Pi}_{vect} \otimes \hat{P} \; )
$
commutes with the wave operator in (25a).
Further, from the equation on proper values $\hat{N} \;
\Phi_{\epsilon jm} = N\; \Phi_{\epsilon jm} $ it follows
$$
N = \delta \; (-1)^{j+1} \; : \; \;
 g_{1} = \delta \;
f_{1} \; , \;\;  g_{2} = \delta \; f_{4} \; , \;\; g_{3} = \delta \;
f_{3} \; , \;\; g_{4} = \delta \; f_{2} \; ,
$$
$$
g_{5} = \delta \;
f_{7} \; , \;\; g_{6} = \delta \; f_{6} \; , \;\; g_{7} = \delta \; f_{5}
\; , \;\; g_{8} = - \delta \; f_{10} \; , \;\; g_{9} = -\delta \;
f_{9} \; , \;\; g_{10} = -\delta \; f_{8} \; .
\eqno(28)
 $$

\noindent Finally, it is readily verified that those limitations (28) are
consistent with the above system (27); so we get 10 equations (one
ought to take into account the  relation $c^{\pm} = d^{\mp}$)
$$
-(\; {d \over {dr}} \; + \; {2\over
r}\; )\; f_{6} \; - \; \sqrt{2} \; {1\over r} \; (\; c^{+} \; f_{5}
\; + \; d^{+} \; f_{7}\; )\; \; - \; m \; f_{1} - \;\delta\; W \;
f_{7} = 0 \; ,
$$
$$
i \epsilon \; f_{5} \; + \; i (\; {d \over{dr}}
\; + \; {1\over r}\; ) \; f_{8} \; + \; i \sqrt{2} \; {c^{+} \over r}
\; f_{9} \; - \; m \; f_{2} = 0  \;   ,
$$
$$
i \epsilon \; f_{6} \;
+ \; {2i\over r}\; (\; -c^{+} \; f_{8} \; + \; d^{+} \; f_{10}\;) \;
-\;  m f_{3} \; - \; i \delta \; W \; f_{10} = 0  \; ,
$$
$$
i\epsilon \; f_{7} \;  - \; i (\; {d \over{dr}} \; + \; {1 \over r}\; )\;
f_{10} \;  -  \; i \sqrt{2} \; {d^{+} \over r} \; f_{9} \; -
\; m \; f_{4} \;+ \; i\delta \; W \; f_{9} = 0 \; ,
$$
$$
i\epsilon \; g_{5} \;  + \; i (\; {d \over{dr}} \; + \; {1 \over r}\; )\;
 g_{8} \;  +  \; i \sqrt{2} \; {c^{-} \over r} \; g_{9} \; -
\; m \; g_{2} \;+ \; i  W \; f_{9} = 0 \; ,
$$
$$
i \epsilon \; f_{2} \; + \; \sqrt{2} \; {c{+} \over r}\;  f_{1} \; -
\; m \; f_{5} = 0 \; ,
$$
$$
-i \epsilon \; f_{3} \; - \; {d \over{dr}} \; f_{1} \; -  \; m \; f_{6} = 0
\; ,
$$
$$
i \epsilon \; f_{4} \; + \; \sqrt{2} \; {d^{+} \over r} \; f_{1} \; -
\; m \; f_{7} \; + \;\delta \; W \; f_{1} = 0  \; ,
$$
$$
-i\; (\; {d\over {dr}}\; + \; {1\over r}\; )\; f_{2} \; - \;
i \sqrt{2}\; {c^{+} \over r} \; f_{3} \; - \; m \; f_{8} = 0  \; ,
$$
$$
i \sqrt{2}\; {1\over r} \; (\; c^{+} \; f_{2} \; - \; d^{+} \; f_{4}\; ) \; -
 \; m \;f_{9} \; + \;i\delta \; W \; f_{4} = 0  \; ,
$$
$$
i\; (\; {d \over {dr}\; } \; + \; {1\over r}\; )\; f_{4} \; + \;
i\sqrt{2}\; {d^{+} \over r}\; f_{3} \; - \; m \; f_{10} \; +
\; i \delta W \; f_{3} = 0    \; .
\eqno(29)
$$

\noindent
It is no difficulty to see that this discrete operator is
self-conjugated one, and the relevant  selection  rules on the
composite $N$-parity  are quite available.

\subsection*{7. Discussion}

In author's opinion, analysis of all unusual selection rules with
respect to the quantum number of the generalised momentum $j$ on the
monopole background (studied in the literature), which certainly
exhibit definite traces and accompanying features of the
monopole-based $P$-violation,  accomplishes almost nothing about
quite symmetrical character of that $P$-violation:
$$
\Psi^{\pm eg}(-\vec{r}) = Matrix \;  \Psi^{\mp eg}(\vec{r})  \; .
\eqno(30)
$$

\noindent  Instead, those selection rules rather  agree
passively with the absence of $P$-symmetry in presence of the Abelian
monopole.   In that context,
the task was to clarify the all significance   and
implications of the relation (30) and also
to find the  points where it will play a part (really substantial in the
sense of its experimental and theoretical manifestations).

The present study has shown that the general outlook on
this matter which prescribes to consider a magnetic charge as
pseudo-scalar under $P$-reflection  seem hardly effective one as we turn
to the most reliable matter --- relevant  selection rules.
In author's opinion, the assertion that the magnetic charge $g$ is a
pseudo-scalar provides rather accidental (though reasonable at first
glance) interpretation of the information carried by the
relation (30).   In any case, the non-existence  of the relevant
selection rules needs to
be understood and rationalised in term of firmly established and
reliable principles.
In that sense, the main suggestion of the paper
--- to formulate some weak points  of this (pseudo scalar) line of
arguments in terms of the property of non-self-conjugacy seemingly
supplies a firm mathematical base for their  discussion.
Because of that non-self-conjugacy, the pseudo scalar nature
of the magnetic charge should be used in theoretical constructions
with extreme caution so as not to lead us   to quite speculative
results.

The analysis above also has shown a contrasting relationship
between Abelian and non-Abelian models regarding the monopole
$P$-(a)symmetry. It may be noticed that just those mathematical
relations which supply ``bad'' peculiarities in the~Abelian
theory  produce, under other circumstances, ``good''
characteristics of the corresponding non-Abelian problems.

\subsection*{Acknowlegements}

In conclusion. I would like to express my gratitude to Professor
A.~A.~Bogush, who looked
through the manuscript and contributed many suggestions for correction and
addition. Also, I am indebted to Dr V.~V.~Gilewsky for his wholehearted
support.

\end{document}